\documentclass[
  journal=pasa,
  manuscript=letter,
  year=2026,
  volume=XX,
]{cup-journal}

\usepackage{amsmath}
\usepackage[nopatch]{microtype}
\usepackage{booktabs}

\usepackage{xspace}
\usepackage[nolist]{acronym}
\usepackage{multicol}
\usepackage{supertabular}
\usepackage{multirow}
\usepackage{comment}

\usepackage{microtype,siunitx,booktabs}
\usepackage{makecell}
\usepackage{hyperref} 
\hypersetup{colorlinks,citecolor=blue,linkcolor=blue,urlcolor=blue}
\usepackage[nolist]{acronym}
\usepackage[version=3]{mhchem}
\usepackage{mathtools}
\usepackage{adjustbox}
\usepackage[normalem]{ulem}
\usepackage{xcolor}

\usepackage{comment}
\usepackage{amssymb}
\usepackage{upgreek}
\usepackage{amsfonts}
\usepackage{pifont} % for xmark
\usepackage{tikz} % for checkmark
\usepackage{graphicx}
\usepackage{float}
\usepackage{footmisc}

\begin{acronym}[AWGN]
\acro{2MASS}{Two Micron All Sky Survey}
\acro{2MASX}{Two Micron All Sky Survey Extended Source Catalogue}
\acro{30Dor}[30 Dor]{30~Doradus}
\acro{AeReS}{Aegean Residual}
\acro{AGN}{active galactic nucleus}
\acrodefplural{AGN}{active galactic nuclei}
\acro{ATCA}{Australia Telescope Compact Array}
\acro{ATESP}{Australia Telescope ESO Slice Project}
\acro{ATNF}{Australia Telescope National Facility}
\acro{ATOA}{Australia Telescope Online Archive}
\acro{AT20G}{Australia Telescope 20 GHz Survey}
\acro{ASKAP}{Australian Square Kilometre Array Pathfinder}
\acro{ARC}{Australian Research Council}
\acro{BETA}{Boolardy Engineering Test Array}
\acro{BGG}{Brightest Galaxy Group}
\acro{BL}[BL Lac]{BL Lacertae Objects}
\acro{CASS}{CSIRO Astronomy and Space Science}
\acro{CABB}{Compact Array Broadband Back-end}
\acro{CHII}[{\sc CHii}]{Compact \textsc{Hii}}
\acro{CMB}{Cosmic Microwave Background}
\acro{CSIRO}{Australian Commonwealth Scientific and Industrial Research Organisation}
\acro{CSM}{circumstellar material}
\acro{CSS}{Compact Steep Spectrum}
\acro{DD}{double degenerate}
\acro{DECaLS}{Dark Energy Camera Legacy Survey}
\acro{DESI}{Dark Energy Spectroscopic Instrument}
\acro{DR3}{Data Release 3}
\acro{DS9}[\textsc{DS9}]{\textsc{SAOImage DS9}}
\acro{DSA}{diffusive shock acceleration}
\acro{DSS}{Digital Sky Survey}
\acro{DSS2}{Digital Sky Survey 2}
\acro{EBHIS}{Effelsberg--Bonn H{\sc i} Survey}
\acro{EM}{Electromagnetic}
\acro{EMU}{Evolutionary Map of the Universe}
\acro{eROSITA}{extended R\"Ontgen Survey with an Imaging Telescope Array}
\acro{ev}[eV]{electronvolt\acroextra{: 1 eV $\approx 1.6 \times 10^{-19}$ J}}
\acro{FITS}[\textsc{Fits}]{Flexible Image Transport System}
\acro{FRB}{Fast Radio Bursts}
\acro{FSRQ}{Flat Spectrum Radio Quasars}
\acro{FWHM}{Full Width at Half-Maximum}
\acro{GASS}{Galactic All-Sky Survey}
\acro{GB}{giant branch}
\acro{GLARE}{Galaxies with Large-scale Ambient Radio
Emission}
\acro{GLEAM}{GaLactic and Extragalactic All-sky MWA Survey}
\acro{GMRT}{Giant Metrewave Radio Telescope}
\acro{GRB}{Gamma-ray burst}
\acro{GPS}{Gigahertz Peak Spectrum}
\acro{HASH}{Hong Kong/AAO/Strasbourg H$\alpha$}
\acro{HEMT}{High Electron Mobility Transistor}
\acro{H.E.S.S.}{High Energy Stereoscopic System}
\acro{HFP}[HFP]{High-Frequency Peaker}
\acro{HPBW}{Half Power Beam Width} 
\acro{HzRGs}{High Redshift Radio Galaxies}
\acro{pHFP}[pHFP]{Potential High Frequency Peaker}
\acro{HI}[H{\sc i}]{Neutral Atomic Hydrogen} 
\acro{HST}{\textit{Hubble Space Telescope}}
\acro{ICM}{intracluster medium}
\acro{IAU}{International Astronomical Union}
\acro{IFRSs}{Infrared Faint Radio Sources}
\acro{ISM}{Interstellar Medium}
\acro{IFRS}{Infrared Faint Radio Source}
\acro{IR}{infrared}
\acro{JY}[Jy]{Jansky\acroextra{, 1 Jy = $10^{-26} \times \mathrm{W~ m}^{-2}~\mathrm{Hz}^{-1}$}} 
\acro{LLS}{Largest Linear Size}
\acro{LFAA}{Low-Frequency Aperture Array}
\acro{LMC}{Large Magellanic Cloud}
\acro{LOFAR}{Low-Frequency Array}
\acro{LSO}[LSOs]{Large Scale Objects}
\acro{LBV}{Luminous Blue Variable}
\acro{MACHO}{Massive Astrophysical Compact Halo Objects}
\acro{MC}[MC]{Magellanic Cloud}    
\acro{MCELS}{Magellanic Cloud Emission Line Survey}
\acro{MS}{Magellanic Stream}
\acro{MW}{Milky Way}
\acro{MIRIAD}[\textsc{Miriad}]{Multichannel Image Reconstruction, Image Analysis and Display}
\acro{MIT}{Massachusetts Institute of Technology}
\acro{MOST}{Molonglo Observatory Synthesis Telescope}
\acro{MGPS-2}{second epoch Molonglo Galactic Plane Survey}
\acro{MQS}{Magellanic Quasars Survey}
\acro{MRC}{Molonglo Reference Catalogue of Radio Sources}
\acro{MWA}{Murchison Widefield Array}
\acro{NED}{NASA/IPAC Extragalactic Database}
\acro{NRAO}{National Radio Astronomy Observatory}
\acro{NSR}{Nova Super Remnant}
\acro{NVSS}{NRAO VLA Sky Survey}
\acro{OPAL}{Online Proposal Applications \& Links}
\acrodefplural{ORC}{Odd Radio Circles}
\acro{ORC}{Odd Radio Circle}
\acro{OVV}{Optically Violent Variable Quasars}    
\acro{PACS}{Photodetector Array Camera and Spectrometer}
\acro{PAF}{Phased Array Feed}
\acro{pc}{parsec\acroextra{: 1 pc $\simeq 3.09 \times 10^{16}$ m}}
\acro{PM}{proper motion}
\acro{PMN}{Parkes-MIT-NRAO}
\acro{PN}{Planetary Nebula}
\acro{PNe}{Planetary Nebulae}
\acro{PWN}{pulsar-wind nebula}
\acrodefplural{PWN}[PWNe]{pulsar-wind nebulae}
\acro{QSO}{Quasi-Stellar Object}
\acro{RA}{Right Ascension}
\acro{RACS}{Rapid ASKAP Continuum Survey}
\acro{RFI}{Radio-Frequency Interference}
\acro{RMS}[RMS]{Root Mean Squared}
\acro{RM}{rotation measure}
\acro{SARAO}{South African Radio Astronomy Observatory}
\acro{SDSS}{Sloan Digital Sky Survey}
\acro{SED}{spectral energy distribution}
\acro{SGPS}{Southern Galactic Plane Survey}
\acro{SHS}{SuperCOSMOS H$\alpha$ Survey}
\acro{SI}[$\alpha$]{spectral index\acroextra{, $S \propto \nu^\alpha$}}
\acro{SKA}{Square Kilometre Array}
\acro{SMB}{Super Massive Blackholes}
\acro{SMC}{Small Magellanic Cloud}
\acro{SN}{supernova}
\acro{SUMSS}{Sydney University Molonglo Sky Survey}
\acro{TOPCAT}[\textsc{Topcat}]{Tool for OPerations on Catalogues And Tables}
\acro{USS}{Ultra Steep Spectrum}
\acro{YSOs}{young stellar objects}
\acro{WALLABY}{Widefield ASKAP L-band Legacy All-sky Blind surveY}
\acro{WBAC}{Wide-Band Analogue Correlator}
\acro{WD}{white dwarf}
\acro{WIFES}[WiFeS]{Wide-Field Spectrograph}
\acro{WISE}{Wide-Field Infrared Survey Explorer}
\acro{WR}{Wolf-Rayet}
\acro{UGC}{Unresolved Galaxy Classifier}
\acro{VLA}{Very Large Array} 
\acro{VLBI}{Very Long Baseline Interferometry} 
\acro{VLSR}[\textbf{$v_{lsr}$}]{Velocity in the Line of Sight}
\acro{SMASH}{Survey of the Magellanic Stellar History}
\acro{SNR}{supernova remnant}
\acrodefplural{SNR}{supernova remnants}
\acro{SD}{single-degenerate}
\acro{SUMSS}{Sydney University Molonglo Sky Survey}
\acro{SMBH}{Super Massive Black Hole}
\acro{POSSUM}{Polarisation Sky Survey of the Universe's Magnetism}
\acro{CASDA}{CSIRO ASKAP Science Data Archive}
\acro{MIPS}{Multiband Imaging Photometer for Spitzer}
\acro{SN}{Supernova}
\acro{CC}{Core Collapse}
\acro{Cas A}{Cassiopeia A}
\acro{CARTA}{Cube Analysis and Rendering Tool for Astronomy}
\acro{IRAC}{Infrared Array Camera}
\acro{MIR}{Mid-Infrared}
\acro{SMGPS}{SARAO MeerKAT Galactic Plane Survey}
\acro{UHE}{Ultra High Energy}
\acro{RN}{Reflection Nebula}
\acro{RNe}{reflection nebulae}
\acro{DR2}{Data Release 2}
\acro{SVM}{support vector machine}
\end{acronym}
\def\wisegal{WISEA~J094409.17$-$751012.8}

\newcommand{\ergcm}[1]{erg\,cm$^{-2}$\,s$^{-1}$}

\newcommand{\D}{$^\circ$}

\newcommand{\kms}{km\,s$^{-1}$}

\newcommand{\farcs}{\mbox{\ensuremath{.\!\!^{\prime\prime}}}}%  % fractional arcsecond symbol: 0.''0
\def\arcmin{\hbox{$^\prime$}}
\def\arcsec{\hbox{$^{\prime\prime}$}}
\def\kms{km\,s$^{-1}$}

\title{ASKAP EMU detection of an Odd Radio Circle (ORC) candidate: J094412$-$751016 (Anglerfish)}

\author{M. D. Filipovi\'c}
\affiliation{Western Sydney University, Locked Bag 1797, Penrith South DC, NSW 2751, Australia}
\email[M. D. Filipovi\'c]{m.filipovic@westernsydney.edu.au}

\author{Z. J. Smeaton}
\affiliation{Western Sydney University, Locked Bag 1797, Penrith South DC, NSW 2751, Australia}

\author{A. C. Bradley}
\affiliation{Western Sydney University, Locked Bag 1797, Penrith South DC, NSW 2751, Australia}

\author{R. Kothes}
\affiliation{Dominion Radio Astrophysical Observatory, Herzberg Astronomy \& Astrophysics, National Research Council Canada, P.O. Box 248, Penticton, BC V2A6J9, Canada}

\author{E. J. Crawford}
\affiliation{Western Sydney University, Locked Bag 1797, Penrith South DC, NSW 2751, Australia}

\author{A. Ahmad}
\affiliation{Western Sydney University, Locked Bag 1797, Penrith South DC, NSW 2751, Australia}

\author{T. Akahori}
\affiliation{Mizusawa VLBI Observatory, National Astronomical Observatory Japan, 2-21-1 Osawa, Mitaka, Tokyo 181-8588, Japan}

\author{L. Barnes}
\affiliation{Western Sydney University, Locked Bag 1797, Penrith South DC, NSW 2751, Australia}

\author{C. Bordiu}
\affiliation{INAF-Osservatorio Astrofisico di Catania, Via Santa Sofía 78, I-95123 Catania, Italy}

\author{S. Dai}
\affiliation{Australia Telescope National Facility, CSIRO, Space and Astronomy, PO Box 76, Epping, NSW 1710, Australia}
\alsoaffiliation{Western Sydney University, Locked Bag 1797, Penrith South DC, NSW 2751, Australia}

\author{S. Duchesne}
\affiliation{Australia Telescope National Facility, CSIRO Space and Astronomy, PO Box 1130, Bentley, WA 6151, Australia}

\author{Y. A. Gordon}
\affiliation{Department of Physics, University of Wisconsin-Madison, 1150 University Avenue, Madison, WI 53706, USA}

\author{N. Gupta}
\affiliation{Australia Telescope National Facility, CSIRO Space and Astronomy, PO Box 1130, Bentley, WA 6151, Australia}

\author{A. M. Hopkins}
\affiliation{School of Mathematical and Physical Sciences, 12 Wally's Walk, Macquarie University, NSW 2109, Australia}

\author{B. S. Koribalski}
\affiliation{Australia Telescope National Facility, CSIRO, Space and Astronomy, PO Box 76, Epping, NSW 1710, Australia}
\alsoaffiliation{Western Sydney University, Locked Bag 1797, Penrith South DC, NSW 2751, Australia}

\author{S. Lazarevi\'c}
\affiliation{Western Sydney University, Locked Bag 1797, Penrith South DC, NSW 2751, Australia}
\alsoaffiliation{Australia Telescope National Facility, CSIRO, Space and Astronomy, PO Box 76, Epping, NSW 1710, Australia}
\alsoaffiliation{Astronomical Observatory, Volgina 7, 11060 Belgrade, Serbia}

\author{D. Leahy}
\affiliation{Department of Physics and Astronomy, University of Calgary, Calgary, Alberta, T2N IN4, Canada}

\author{K. J. Luken}
\affiliation{Western Sydney University, Locked Bag 1797, Penrith South DC, NSW 2751, Australia}

\author{P. J. Macgregor}
\affiliation{Western Sydney University, Locked Bag 1797, Penrith South DC, NSW 2751, Australia}

\author{A. Mailvaganam}
\affiliation{School of Mathematical and Physical Sciences, 12 Wally's Walk, Macquarie University, NSW 2109, Australia}

\author{S. Mehmood}
\affiliation{Western Sydney University, Locked Bag 1797, Penrith South DC, NSW 2751, Australia}

\author{R. P. Norris}
\affiliation{Western Sydney University, Locked Bag 1797, Penrith South DC, NSW 2751, Australia}

\author{N. Novaretti}
\affiliation{Western Sydney University, Locked Bag 1797, Penrith South DC, NSW 2751, Australia}

\author{L. A. F. Park}
\affiliation{Western Sydney University, Locked Bag 1797, Penrith South DC, NSW 2751, Australia}

\author{S. Riggi}
\affiliation{INAF-Osservatorio Astrofisico di Catania, Via Santa Sofía 78, I-95123 Catania, Italy}

\author{C.~J.~Riseley}
\affiliation{Ruhr University Bochum, Faculty of Physics and Astronomy, Astronomical Institute (AIRUB), Universitätsstraße 150, 44801 Bochum, Germany}

\author{G. Rowell}
\affiliation{School of Physics, Chemistry and Earth Sciences, The University of Adelaide, Adelaide 5005, Australia}

\author{M. Sasaki}
\affiliation{Dr. Karl Remeis Observatory, Erlangen Centre for Astroparticle Physics, Friedrich-Alexander-Universit\"{a}t Erlangen-N\"{u}rnberg, Sternwartstra{\ss}e 7, 96049 Bamberg, Germany}

\author{S. S. Shabala}
\affiliation{School of Natural Sciences, Private Bag 37, University of Tasmania, Hobart, TAS 7001, Australia}

\author{S. Taziaux}
\affiliation{Ruhr University Bochum, Faculty of Physics and Astronomy, Astronomical Institute (AIRUB), Universitätsstraße 150, 44801 Bochum, Germany}

\author{N. F. H. Tothill}
\affiliation{Western Sydney University, Locked Bag 1797, Penrith South DC, NSW 2751, Australia}

\author{D. Uro\v{s}evi\'c}
\affiliation{Department of Astronomy, Faculty of Mathematics, University of Belgrade, Studentski trg 16, 11000 Belgrade, Serbia}

\author{V. Velovi\'c}
\affiliation{Western Sydney University, Locked Bag 1797, Penrith South DC, NSW 2751, Australia}

\author{T. Vernstrom}
\affiliation{Australia Telescope National Facility, CSIRO Space and Astronomy, PO Box 1130, Bentley, WA 6151, Australia}
\alsoaffiliation{International Centre for Radio Astronomy Research (ICRAR), The University of Western Australia, 35 Stirling Highway}

\author{J. L. West}
\affiliation{Department of Physics and Astronomy, University of Calgary, Calgary, Alberta, T2N IN4, Canada}

\author{T. Zafar}
\affiliation{School of Mathematical and Physical Sciences, 12 Wally's Walk, Macquarie University, NSW 2109, Australia}

 %% First letter not capped

\begin{document}

\begin{abstract}

We report diffuse extended radio-continuum emission spatially coinciding with the IR source \wisegal, and a semi-variable star, V687~Carinae.
%in the field of view. 
We use 944\,MHz radio data from the large-scale \ac{EMU} survey to analyse this diffuse emission (EMU~J094412$-$751016), which we nickname ``Anglerfish''. We investigate if the spatially correlated \ac{IR} source, \wisegal, is physically related to Anglerfish. The \ac{IR} colours of \wisegal\ are indicative of an elliptical galaxy, raising the possibility that Anglerfish may belong to the newly-discovered class of extragalactic radio sources known as \acp{ORC} with \wisegal\ as the host galaxy. We also investigate the possibility that Anglerfish is physically related to the star, V687~Carinae, and whether it may be a remnant from a previous epoch of stellar mass-loss. We determine that a physical association between the radio emission and the star is unlikely due to the %emission's non-thermal nature and the 
star's weak stellar winds compared to the theoretical expansion velocity of the `shell'. It is possible that Anglerfish may be a Galactic high-latitude \ac{SNR}; however, we find that the observed size and luminosity are not consistent with this scenario. We also investigate the \ac{ORC} scenario, which we deem the most likely scenario based on the Anglerfish's observed properties such as size, brightness, lack of other frequency detections, and possible host galaxy identification. We therefore propose Anglerfish as an \ac{ORC} candidate, but note that additional radio and optical observations are vital to further constrain the properties and confirm this classification. %understanding and resolving the mystery of Anglerfish's nature.

\end{abstract}

\acresetall

\section{Introduction}
\label{Section:Introduction}

The \ac{EMU} \citep{Norris2021,2025PASA...42...71H} survey is a large-scale radio survey currently being undertaken by the \ac{ASKAP}~\citep{2021PASA...38....9H} telescope to map the southern sky at 944\,MHz. The improved resolution and sensitivity of modern observatories (e.g. \ac{ASKAP}, MeerKAT \citep{2016mks..confE...1J}, etc), allow the discovery and analysis of previously unseen radio objects and emissions. These facilities have revealed numerous new \acp{SNR} \citep{2022MNRAS.512..265F,2023AJ....166..149F,2024PASA...41..112F,2025PASA...42..104F,2025A&A...693L..15S}, and \ac{SNR} candidates \citep{2024MNRAS.534.2918S,2024RNAAS...8..158S,2024RNAAS...8..107L}. They have also uncovered several other objects of interest, including \acp{PWN} \citep{2024PASA...41...32L,2025MNRAS.537.2868A}; \ac{RNe} \citep{2025PASA...42...32B}; \acp{AGN} \citep{2023MNRAS.523.1933V, 2022MNRAS.516.1865V}; and enigmatic objects of unknown origin \citep{2024A&A...690A..53B,Stingrays}. \ac{ASKAP}, alongside other instruments such as MeerKAT, \ac{GMRT}~\citep{2001aprs.conf..237A} and \ac{LOFAR}~\citep{2013A&A...556A...2V},  have also been instrumental to the search for radio counterparts of neutrino detections~\citep{Filipović_2025}, as well as the discovery of \acp{ORC}~\citep{Norris2021ORC,2022MNRAS.513.1300N,10.1093/mnrasl/slab041,2025MNRAS.543.1048H,2026arXiv260218294D}. These \acp{ORC} are an interesting class of new radio objects, appearing exclusively at radio-continuum frequencies. They typically display a circular structure, generally a few arcminutes in size, and often have a central elliptical galaxy visible in the optical, which may be the host of the emission. These properties can vary across different \acp{ORC}, however, and the origin of these objects is still being investigated~\citep{Norris2021ORC}.

%%%%%%%%%%%%%%%%%%%%%%%%%%%%%%%%%%%%%%%%%%%% Fig. 1 %%%%%%%%%%%%%%%%%%%%%%%%%%%%%%%%%%%%%%%%%%%%%%%%%%%%%%%%%%%%%%%%%%%%%%%%%%%%%%%%%%%%
   \begin{figure*}[!ht]
\centering
\vskip-2mm
    \includegraphics[width=\linewidth]{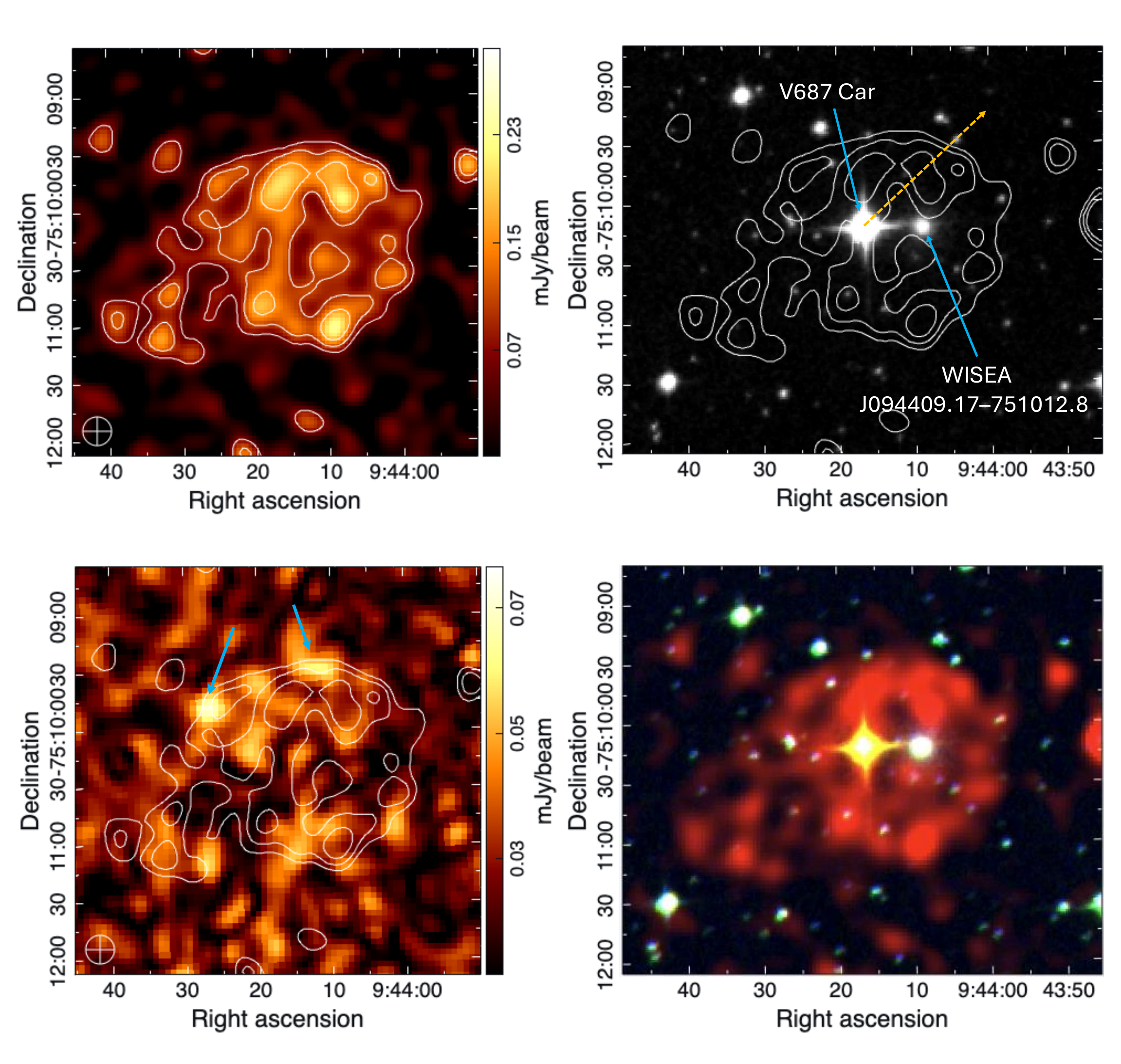}
    \vskip-1mm
    \caption{Four-panel image of Anglerfish radio-continuum emission. {\bf Top left: } 944\,MHz \ac{ASKAP} radio-continuum image (linearly scaled) with a measured \ac{RMS} noise level of $\sim$25-30$\mu$Jy beam$^{-1}$, and a 15\arcsec$\times$15\arcsec convolved beam size shown in the bottom left corner. Contours are from the same image at levels of 60, 100, and 150\,$\mu$Jy beam$^{-1}$. {\bf Top right: } DSS2~IR image. The variable star V687~Car and the elliptical galaxy \wisegal\ are annotated in the image with the solid blue arrows. The dashed orange arrow shows the direction of proper motion of V687~Car. The image is linearly scaled and the contours are from the radio-continuum image at the same levels as the top left panel. {\bf Bottom left: }Polarised intensity (PI) image with an \ac{RMS} noise level of 10\,$\mu$Jy\,beam$^{-1}$. The image is linearly scaled, and the contours are from the radio-continuum image at the same levels as the top left panel. The image is convolved to a beam size of 18\arcsec$\times$18\arcsec, shown in the bottom left corner. There are two point sources in PI at levels of $\sim$8$\sigma$ and $\sim$6$\sigma$, indicated by the blue arrows. {\bf Bottom right: }RGBY image using radio, optical, and IR data. Red is EMU 944\,MHz, green is DSS2~Red, blue is DSS2~blue, and yellow is DSS2~IR. All images are linearly scaled.  
    }
    \label{Figure:RGBY}
\end{figure*}
%%%%%%%%%%%%%%%%%%%%%%%%%%%%%%%%%%%%%%%%%%%%%%%%%%%%%%%%%%%%%%%%%%%%%%%%%%%%%%%%%%%%%%%%%%%%%%%%%%%%%%%%%%%%%%%%%%%%%%%%%%%%%%%%%%%%%%%%%

One of the most recent \ac{EMU} survey datasets reveals a patch of diffuse radio emission, EMU~J094412$-$751016, which we nickname Anglerfish\footnote{Due to the apparent radio morphology being reminiscent of the shape of such a fish (the Anglerfish are ray-finned fish in the order Lophiiformes.)}. This is a distinct extended emission source, composed of an $\sim$55\arcsec\ radius circular component, and an $\sim$30\arcsec\ extension to the south-east. We find the emission is spatially correlated with two obvious optical and \ac{IR} objects (see Figure~\ref{Figure:RGBY}). We identify the first object 
%located at RA(J2000)=09$^{\rm h}$44$^{\rm m}$09.2$^{\rm s}$, Dec(J2000)=$-$75$^\circ$10$'$13\farcs2 
as a known IR source \wisegal, and a second object 
%located at RA(J2000)=09$^{\rm h}$44$^{\rm m}$16.6$^{\rm s}$, Dec(J2000)=$-$75$^\circ$10$'$12\farcs5 
as the semi-variable star V687~Carinae (hereafter referred to as V687~Car).
%We analyse the IR colours of \wisegal\ and determine it to be an elliptical type galaxy (see Sec.~\ref{ORC candidate}). 

%Here, we present the discovery of the new radio source, Anglerfish. 
In Section~\ref{sec:app obs data} we discuss the data used in this analysis. In Section~\ref{results and discussion} we present our results. In Section~\ref{discussion} we discuss three main origin scenarios for the radio-continuum emission: 1) a remnant of a stellar mass-loss episode from V687~Car, 2) a high-latitude \ac{SNR}, and 3) an \ac{ORC} with \wisegal\ as the host elliptical galaxy. In Section~\ref{Section:Conclusion}, we present our conclusions.

\section{Data}
\label{sec:app obs data}

\ac{ASKAP} observed Anglerfish on 25$^{\rm th}$~March~2025 as part of the \ac{EMU} survey in tile EMU\_0941$-$75 (SB72176). The data were reduced using the standard ASKAPSoft data reduction pipeline, which consists of multi-frequency synthesis imaging, multi-scale cleaning, and self-calibration, followed by convolution to a common beam size of B.S.=15\arcsec$\times$15\arcsec~\citep{2019ascl.soft12003G}. The final 944\,MHz Stokes~$I$ radio image is shown in Figure~\ref{Figure:RGBY}, 
%with a synthesised beam of 15\arcsec$\times$15\arcsec\ and 
and we estimate a local \ac{RMS} noise sensitivity of $\sim$25-30\,$\mu$Jy\,beam$^{-1}$.
%We also analysed the available Stokes~$V$ data of the same observation, but detected no signs of circular polarisation. 
%Finally, we also analysed the available Taylor~1 (T1) image of the observations in order to attempt to estimate an \ac{ASKAP} in-band \ac{SI}. Due to the low surface brightness of the Anglerfish's emission, however, the spectral index results had significant errors, and we were not able to generate an accurate spectral index.

We also use radio polarisation data from the \ac{ASKAP} \ac{POSSUM} survey \citep{2025arXiv250508272G}. These data are obtained from the same observations and scheduling blocks as the \ac{EMU} data. \ac{POSSUM} data consist of Stokes~$Q$ and $U$ polarisation frequency cubes with 1\,MHz channels. These images are used to calculate the polarised intensity (PI), and the resulting polarisation images have a lower resolution than \ac{EMU} (B.S.$=$ 18\arcsec$\times$18\arcsec; see Figure~\ref{Figure:RGBY}, bottom left). %The Stokes~$I$ image is then convolved to the \ac{EMU} resolution to ensure consistency during comparison (see Figure~\ref{Figure:RGBY}).

We use \ac{PM} measurements obtained from the \textit{Gaia} telescope \citep{2016A&A...595A...1G} to determine movements of the star V687~Car. These \ac{PM} values were obtained from \textit{Gaia} \ac{DR3} \citep{2023A&A...674A...1G}. We also use the distance measurement from \citep{2018AJ....156...58B}, derived from the \textit{Gaia} \ac{DR2} parallax (See Section \ref{stellar mass loss}).

%\ac{DR3} provides the most up-to-date and accurate astrometry measurements with which we can derive distance, direction, and velocity for the star.

We use optical data from the \ac{DSS2} optical survey~\citep{1996ASPC..101...88L}, including red, blue, and \ac{IR} bands to analyse the optical properties of the emission, \wisegal, and V687~Car (see Figure~\ref{Figure:RGBY}). 
We also use \ac{WISE}~\citep{2010AJ....140.1868W} \ac{IR} observations from~\citep{2013wise.rept....1C}, specifically the W1 (3.4\,$\mu$m), W2 (4.6\,$\mu$m), and W3 (12\,$\mu$m) bands, to analyse the \ac{IR} colours of \wisegal.

Finally, we searched for any sign of Anglerfish at other frequencies but found no corresponding emission. Specifically, we searched in FIR ({\it Spitzer}), UV ({\it GALEX}), X-ray (RASS, eRASS~DR1, {\it XMM-Newton}, and {\it Chandra}), and $\gamma$-ray (Fermi). Other available radio surveys, including the \ac{RACS}~\citep{2020PASA...37...48M}, \ac{SUMSS}~\citep{2003MNRAS.342.1117M} and \ac{PMN}~\citep{1996ApJS..103..145W} were also searched, but no traces of emission associated with Anglerfish were found.

\section{Results}
\label{results and discussion}

%We detect an extended radio emission object in our \ac{EMU} radio-continuum image (Fig.~\ref{Figure:RGBY}) named EMU~J094412$-$751016 (Anglerfish). 
The Anglerfish radio emission has two distinct components (Figure~\ref{Figure:RGBY}); a circular region centred at RA(J2000) = 09$^{\rm h}$44$^{\rm m}$12.3$^{\rm s}$, Dec(J2000) = $-$75$^\circ$10$'$16\farcs9 (Galactic coordinates: $l$\,=\,291.7\D, $b$\,=\,$-$16.5\D) with radius $\sim$55\arcsec, and a region which extends $\sim$30\arcsec\ towards the south-east. The entire area is elliptical, centred at RA(J2000) = 09$^{\rm h}$44$^{\rm m}$18.6$^{\rm s}$, Dec(J2000)\,=\,$-$75$^\circ$10$'$26\farcs7 (angled at 22 degrees) with semi-axes of 60\arcsec\ and 90\arcsec, which is shown in Figure~\ref{Figure:Measurement}. There are no obvious radio point sources visible within the emission.

\begin{figure}[!ht]
\centering
\vskip-2mm

    \includegraphics[bb= 100 50 360 300,scale=0.95]{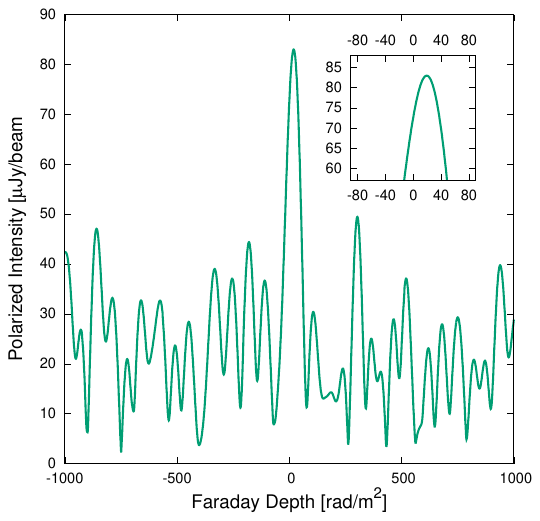}
    \vskip-1mm
    \caption{Faraday Depth Spectrum of the brightest peak in the PI map displayed in Figure~\ref{Figure:RGBY}, where it is indicated by the left blue arrow. In the top right inset, we show an inset zooming in on the peak, indicating a non-zero RM.}
    \label{Figure:FDspec}
\end{figure}

We analysed the Stokes~$V$ \ac{EMU} data, but detected no circular polarisation. We follow a process similar to that used in \citet{Filipović_2025}, using the \ac{POSSUM} data to create Faraday spectra and calculate the PI of Anglerfish. We then calculate the rotation measure of the polarisation, using the rotation measure synthesis technique (see \citealt{Burn_1966,Brentjens_deBruyn_2005} or additionally \citealt{2010ApJ...712.1157H}) as detailed in \citet{Ball2023}, to de-rotate the linear polarisations. The detailed polarisation images are shown in Figure~\ref{Figure:RGBY}, and the PI image achieves an \ac{RMS} noise level of 10\,$\mu$Jy\,beam$^{-1}$, which is taken from the Faraday depth spectra. 
The peak in PI is 83\,$\mu$Jy\,beam$^{-1}$, as can be seen in Figure~\ref{Figure:FDspec} while in total power, we measure 110\,$\mu$Jy\,beam$^{-1}$ in the same location. That gives 75\% of fractional polarisation, which is acceptable, given the above uncertainties.

We observed two areas of polarisation emission, with peaks of 60 and 80\,$\mu$Jy, giving them significance levels of 6 and 8$\sigma$ , respectively (see Figures~\ref{Figure:RGBY} and \ref{Figure:FDspec}). It should be noted that the noise in the Faraday depth spectra is Ricean and not Gaussian, and so we have adopted an 8$\sigma$ detection threshold. This 8$\sigma$ threshold approximately corresponds to a false positive rate of $\sim$4\%~\citep{2012PASA...29..214G}. Therefore, we take the 8$\sigma$ peak as a detection and the 6$\sigma$ peak as a marginal detection. These regions are located on the edge of the Anglerfish's emission, but do not spatially coincide with the brightest Stokes~$I$ emission. These faint polarised sources are consistent with expansion-related polarisation of the shell. If the emission is of synchrotron origin, then this would require coincident magnetic fields. Thus, the fact that the polarisation peaks are anti-correlated with the Stokes~I total intensity implies that 
%The likely non-thermal nature of the emission requires magnetic fields to coincide with the emission. Thus, the polarisation not being associated with the Stokes~I total intensity implies that 
depolarisation may be taking place, in the case of non-thermal emission. This would likely be due to denser or more magnetised media associated with the peak positions of the total intensity. The thermal vs. non-thermal nature of the radio emission is currently unclear, but the current data indicate that non-thermal emission is more likely, as discussed later in this section. The 8$\sigma$ peak to the left in Figure~\ref{Figure:RGBY} is considered significant at the detection threshold (see Figure~\ref{Figure:FDspec}), however, as it is a point-like source, it could be the polarised emission of an unrelated background source. We find that the polarisation results are fairly weak, and thus determine that they should not be relied on for further theoretical analysis.

%%%%%%%%%%%%%%%%%%%%%%%%%%%%%%%%%%%%%%%%%%%% Fig. 2 %%%%%%%%%%%%%%%%%%%%%%%%%%%%%%%%%%%%%%%%%%%%%%%%%%%%%%%%%%%%%%%%%%%%%%%%%%%%%%%%%%%%  
%%%%%%%%%%%%%%%%%%%%%%%%%%%%%%%%%%%%%%%%%%%%%%%%%%%%%%%%%%%%%%%%%%%%%%%%%%%%%%%%%%%%%%%%%%%%%%%%%%%%%%%%%%%%%%%%%%%%%%%%%%%%%%%%%%%%%%%%%%

We use the elliptical region defined above to measure the flux density of Anglerfish using the astronomy software \ac{CARTA}~\citep{CARTA_2018}. We subtract the nearby diffuse background flux density, following the process of \citet{2019PASA...36...48H}, and assume a 10\% error, following the process of \citet{Filipovic2022}. We note that Anglerfish is located away from the Galactic Plane ($b$\,=\,$-$16.5\D) and therefore the contribution of this background noise is minimal. We measure an integrated radio flux density of Anglerfish to be $S_{943\,\rm MHz}$=5.0$\pm$0.5\,mJy. We assume an uncertainty of 10\% due to the faintness of the object, similar to the methods applied in other analyses of low surface-brightness, diffuse, radio objects~\citep{Filipovic2022,Stingrays}. As we have no other detections at radio frequencies, we cannot determine the radio spectral index. We attempted to use the \ac{ASKAP} Taylor~1 (T1) image to estimate the \ac{ASKAP} in-band spectral index. However, given the low surface brightness, the results are unrealistic and have very large errors. Therefore, we instead use the available \ac{GLEAM} 200\,MHz radio data from the \ac{MWA} telescope \citep{2017MNRAS.464.1146H} and 1367\,MHz \ac{RACS} data from the \ac{ASKAP} telescope \citep{2020PASA...37...48M} to estimate limits on the spectral index. We do not detect Anglerfish in either survey, so these upper flux density limits provide a possible spectral index range. 
%The \ac{GLEAM} 200\,MHz flux density upper limit is $\sim$18\,mJy while the \ac{RACS} 1365\,MHz \citep{2023PASA...40...34D} estimate of flux density upper limit is $\sim$3\,mJy which suggests $\alpha\sim-0.9\pm0.5$.

Anglerfish would appear as an extended object in the \ac{RACS} data and as a point source in the \ac{MWA} data, thus the limits are calculated slightly differently. Measuring an upper limit of a non-detection can be sensitive to the noise level used, and so a more robust method is chosen for the \ac{RACS} and \ac{MWA} noise levels. We measured the local \ac{RMS} noise by generating a histogram of the pixel values within a region surrounding the location of Anglerfish and fitting a Gaussian function to the negative pixel values, taking the measured $\sigma$ value as the uncertainty (see Figure~\ref{fig:Noise}). This approach avoids contamination from positive source emission and ensures only the local noise is measured. This resulted in values of 0.24\,mJy for \ac{RACS} and 9.75\,mJy for \ac{MWA}.

To estimate the upper limit for a non-detection of an extended object, we estimate the uncertainty (noise level) over the source area and multiply this by 3 for a 3$\sigma$ detection limit. We use the elliptical region of 60\arcsec$\times$90\arcsec for the source area, $\Omega_{\rm source}$, and the \ac{RACS} beam size of 13.1\arcsec$\times$9.3\arcsec to calculate the number of beams covering the source as $N\,=\,\Omega_{\rm source}/\Omega_{beam}$. Each of these beams has an \ac{RMS} noise level of $\sigma_{\rm rms}$ which we estimate as 0.24\,mJy\,beam$^{-1}$ (Figure~\ref{fig:Noise}, left). We assume that the noise in each beam is uncorrelated and thus independent of each other, and so the total uncertainty follows error propagation for $N$ independent measurements with the same uncertainty $\sigma_{\rm rms}$, which gives $\sigma_{\rm total}$\,=\,$\sigma_{\rm rms}\times\sqrt{N}$. Therefore, for a 3$\sigma$ detection limit, the upper \ac{RACS} 1367\,MHz flux density limit is calculated as $S < 3\times\sigma_{\rm rms}\times\sqrt{\Omega_{\rm source}/\Omega_{\rm beam}}$\,=\,4.8\,mJy. 

The method used for this limit estimation assumes that the individual beams are uncorrelated. For radio interferometric imaging, the image pixels are inherently correlated to an extent due to the Fourier transform process. However, for extended sources which cover multiple beams (in our case, Anglerfish covers $\Omega_{\rm source}/\Omega_{\rm beam}\sim44$ beams), this correlation is typically not significant. To test the impact of potential beam-correlation effects, we independently calculate an upper limit using an empirical approach that inherently accounts for pixel correlations in the data. For this, we generate 25 apertures of the source size (60\arcsec$\times$90\arcsec) distributed on the \ac{RACS} image in blank sky surrounding the source. We measure the integrated flux density for each of these apertures and calculate the standard deviation, which gives a value of 1.38\,mJy. Assuming a 3$\sigma$ detection threshold, this gives an upper limit of 4.1\,mJy. This is slightly lower than the previously calculated 4.8\,mJy (within 16\%) and shows that these correlation effects are not significantly biasing the results. We adopt the more conservative value of 4.8\,mJy as the 1367\,MHz flux density \ac{RACS} upper limit for the subsequent analysis.

For the \ac{MWA} image, as the source would appear as a point source were it detectable, the estimation is simpler and the source area is taken to be equivalent to the beam area. Therefore, the above equation simplifies to an upper limit of $S<3\sigma$, where $\sigma$ is the measured local \ac{RMS} noise which we measure as $\sim$9.75\,mJy\,beam$^{-1}$ (Figure~\ref{fig:Noise}, right). Therefore, the \ac{MWA} 200\,MHz upper limit is taken as $\sim$29.3\,mJy.

\begin{figure}[!ht]
\centering
\vskip-2mm
    \includegraphics[width=\linewidth]{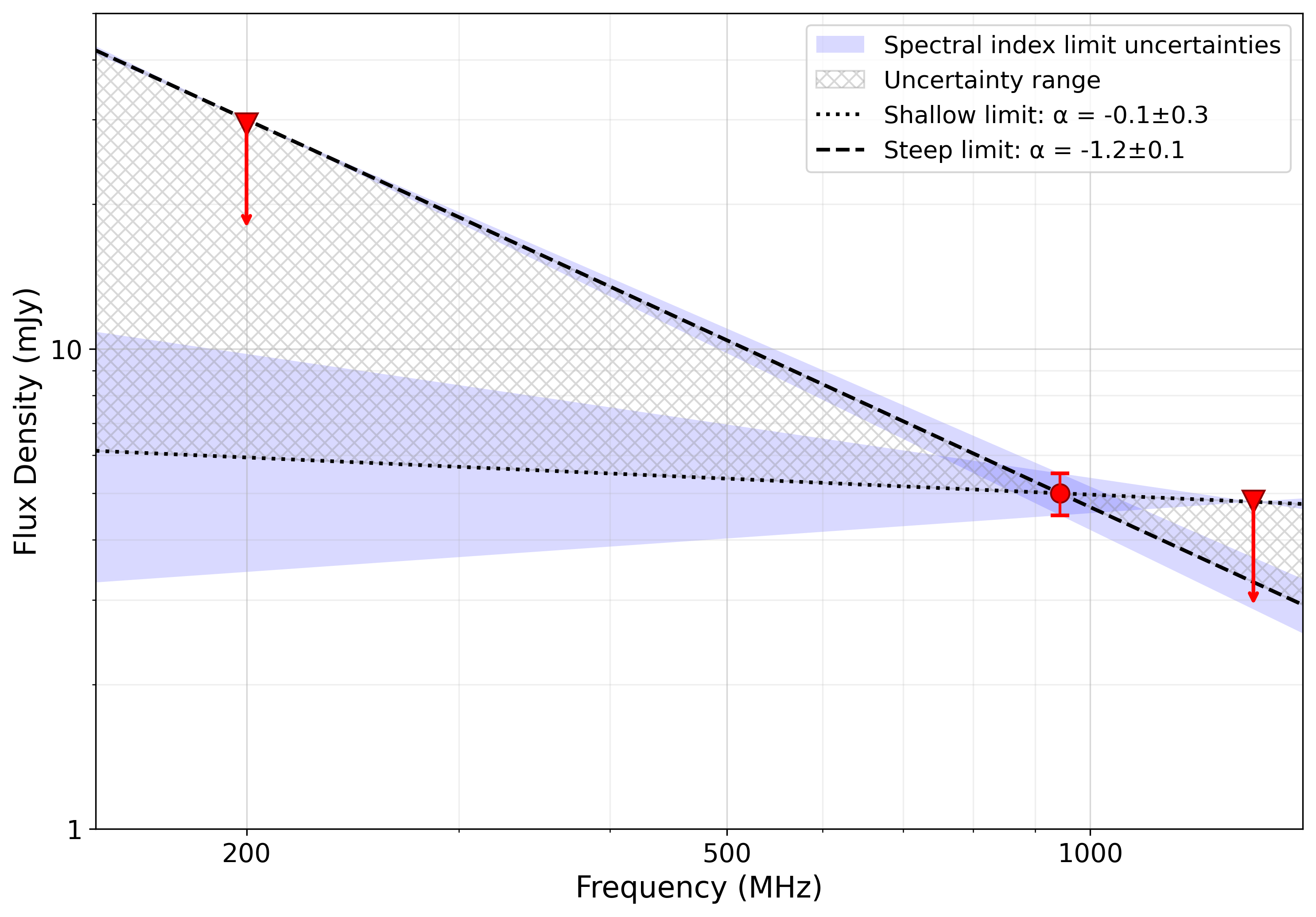}
    \vskip-1mm
    \caption{Spectral index graph of the Anglerfish emission, using the flux density measurement from the \ac{EMU} data, and the upper limits from the \ac{RACS} and \ac{GLEAM} data to generate shallow (dotted line) and steep (dashed line) limits for the spectral index. The uncertainty ranges for each of these spectral index fits is shown as a shaded blue region around each line. The hashed area in between represents the possible spectral index range.
    }
    \label{Figure:Spectral_Index}
\end{figure}

Using the measured 944\,MHz point and the two upper limit flux densities, we calculate a possible spectral index range for the emission. This is done by generating two linear fits as the boundaries of this range, one being through the 944\,MHz \ac{EMU} point and the \ac{RACS} upper limit, and the other through the \ac{EMU} point and the \ac{MWA} upper limit (see Figure~\ref{Figure:Spectral_Index}). We calculate the uncertainties in these limits by also calculating the lines of worst fit through the \ac{EMU} upper and lower uncertainties (5.0$\pm$0.5\,mJy). These spectral index uncertainties are shown as the shaded blue regions in Figure~\ref{Figure:Spectral_Index}. This gives a shallow limit of $\alpha\,=\,-0.1\pm0.3$ and a steep limit of $\alpha\,=\,-1.2\pm0.1$. The data are not sufficient to constrain the spectral index any further, and this entire range (the hashed region in Figure~\ref{Figure:Spectral_Index} of $-1.3<\alpha<0.2$) is the possible spectral index range from the given uncertainties. Both flat and steep spectral indices are thus consistent with the current data, and so both thermal and non-thermal mechanisms are possible. For radio spectral indices, a reasonable cut-off between thermal and non-thermal emission can be assumed to be $-0.3$, with steeper values likely indicating non-thermal, synchrotron emission. Taking $\alpha=-0.3$ as a general dividing line, this shows that two-thirds of the entire uncertainty range is within the non-thermal regime ($\alpha<-0.3$) and one-third is flatter ($\alpha>-0.3$). Therefore, with no other data available, we conclude a non-thermal spectral index is more likely from this range, but emphasise that it cannot be properly constrained without more data. 
%Most of the {\bf spectral index} range covers steeper values, indicating more likely non-thermal emission, but this is not fully constrained without further data.
The current data do not allow us to test the possibilities of any more complex spectral shapes, and a linear fit is assumed.
If the spectral index is indeed non-thermal, this would support the \ac{SNR} and \ac{ORC} scenarios described in Section~\ref{SNR section} and Section~\ref{ORC candidate}. Conversely, a flatter spectral index instead supports the stellar outburst scenario described in Section~\ref{stellar mass loss}. %With the current range available, we deem that the non-thermal scenario is more likely, but emphasise that the spectral index value is only weakly constrained with the current data.

%%%%%%%%%%%%%%%%%%%%%%%%%%%%%%%%%%%%%%%%%%%% Fig. 2 %%%%%%%%%%%%%%%%%%%%%%%%%%%%%%%%%%%%%%%%%%%%%%%%%%%%%%%%%%%%%%%%%%%%
%%%%%%%%%%%%%%%%%%%%%%%%%%%%%%%%%%%%%%%%%%%%%%%%%%%%%%%%%%%%%%%%%%%%%%%%%%%%%%%%%%%%%%%%%%%%%%%%%%%%%%%%%%%%%%%%%%%%%%

We find that Anglerfish emits exclusively at radio frequencies, as demonstrated by the lack of detection at other frequencies across multiple surveys (listed at the end of Section~\ref{sec:app obs data}). We searched for potential counterparts for this emission in the \ac{HASH} \ac{PN} catalogue~\citep{2016JPhCS.728c2008P}, and the Galactic \ac{SNR} catalogue of \citet{2025JApA...46...14G}, but found no corresponding sources. We therefore investigate the two distinctive, centrally positioned optical sources (see Figure~\ref{Figure:RGBY}) as prime candidates for the origin of this radio-continuum emission. The yellowish source on the left is the semi-variable star V687~Car, and the white right-hand point source is the elliptical galaxy \wisegal.

\section{Discussion}
\label{discussion}

\subsection{Anglerfish as stellar (V687~Car) mass loss episode}
\label{stellar mass loss}

%%%%%%%%%%%%%%%%%%%%%%%%%%%%%%%%%%%%%%%%%%%%%%%%%%%%%%%%%%%%%%%%%%%%%%%%%%%%%%%%%%%%%%%%%%%%%%%%%%%%%%%%%%%%%%%%%%%%%%%%%
\begin{figure}
\centering
\vskip-2mm
    \includegraphics[width=\linewidth]{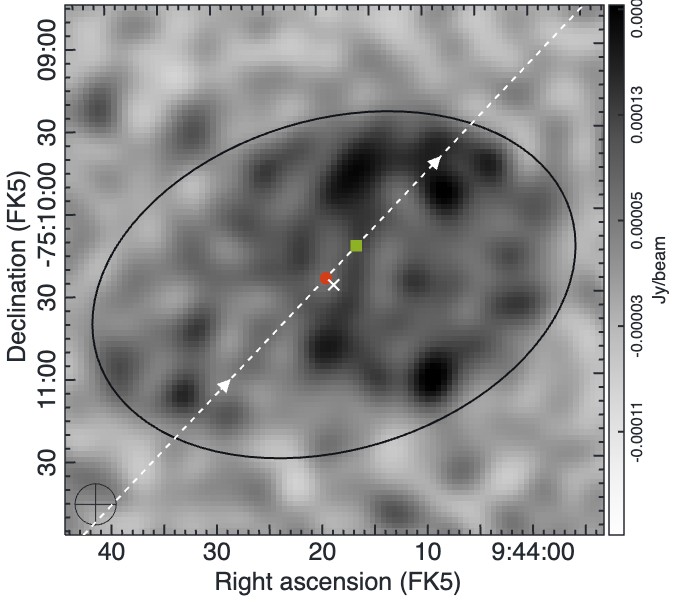}
    \vskip-1mm
    \caption{\ac{ASKAP} \ac{EMU} 944~MHz image of the Anglerfish radio emission with superimposed measurements used in section \ref{stellar mass loss}. The image is linearly scaled with the beam size shown in the bottom left corner. The black circle denotes the elliptical region defined in Section~\ref{results and discussion}. The white dashed line represents V687~Car's tangential movement, with two arrows indicating its direction. The green square shows the location of V687~Car, the white `X' shows the geometric centre of the emission (discussed in section \ref{results and discussion}, and the red circle denotes V687~Car's closest approach to the geometric centre.}
    \label{Figure:Measurement}
\end{figure}
%%%%%%%%%%%%%%%%%%%%%%%%%%%%%%%%%%%%%%%%%%%%%%%%%%%%%%%%%%%%%%%%%%%%%%%%%%%%%%%%%%%%%%%%%%%%%%%%%%%%%%%%%%%%%%%%%%%%%%%%%

V687~Car (also referred to as IRAS~09440$-$7456) is a semi-regular variable star first identified by \cite{2007PZP.....7...20B}. The online SIMBAD database lists V687~Car as a Mira~Ceti type variable, for which it references \citet{2009yCat....102025S}. However, this entry does not list the star as a Mira~Ceti type variable. Subsequent observations have confirmed its variability to be semi-variable type~A~\citep{2017ARep...61...80S}. Despite the star's known variability, there is no clear indication of its spectral classification. \textit{Gaia} \ac{DR3} \citep{2023A&A...674A...1G} lists an effective temperature of 3285.7~K and an absolute magnitude in the G band of -1.33. Tracing these values on the Hertzsprung-Russell diagram \citep{book2} indicates that V687~Car is an M-Type Giant.
%\footnote{The catalogue information is available at the webpage: \url{http://www.sai.msu.su/gcvs/cgi-bin/search2.cgi?search=V687+Car}}

Semi-regular variable stars are evolved giants with intermediate to late spectral types \citep[M, C, S etc.][]{1988A&A...200...85L}). They typically exhibit brightness variations with periods ranging from $\sim$35$-$1200\,days, which can be interspersed by irregular behaviour. Unlike Mira variables, which display large amplitude brightness variations, semi-regular type~A stars have smaller amplitude changes \citep[$<$2.5\,mag; ][]{2021A&A...656A..66T}. V687~Car's variability is measured as fluctuating between apparent V-band magnitudes of $\sim$13$-$14.5 with a period of 240\,days~\citep{2007PZP.....7...20B}, as expected for semi-regular type~A stars. These stars can undergo significant mass loss~\citep{2011ApJ...741...54C}, resulting in weak thermal free-free radio emission from ionised winds or circumstellar envelopes, and in some cases, from molecular maser emission~\citep{2015AJ....149..100M}. \citet{2021AJ....161..147B} estimated a distance to V687~Car of $2.08^{+0.29}_{-0.22}$~kpc, and if we place the radio--continuum feature at this distance, its size would be $\sim$1.21$\times$1.82\,pc (based on the elliptical region defined in Section \ref{results and discussion}). 

V687~Car has FK5 \acp{PM} of $\mu_{\alpha}$:~$-6.877\pm 0.066$\,mas\,yr$^{-1}$ and $\mu_{\delta}$:~$5.374\pm 0.065$\,mas\,yr$^{-1}$. To account for peculiar velocity, we first converted the FK5 \acp{PM} to Galactic Coordinates; $\mu_{l}$:~$-8.699\pm 0.087$\,mas\,yr$^{-1}$ and $\mu_{b}$:~$-0.697\pm 0.069$\,mas\,yr$^{-1}$. Then, using the equations discussed by \citet[][Equations; 2a, 2b, 3a, and 3b]{2007A&A...467L..23C}, we derive peculiar \acp{PM} of V687~Car; $\mu_{l}$:~$-3.149\pm 0.087$\,mas\,yr$^{-1}$ and $\mu_{b}$:~$0.178\pm 0.069$\,mas\,yr$^{-1}$. These correspond to peculiar velocities of 31.1$\pm$0.9\,km\,s$^{-1}$ and 1.8$\pm$0.7\,km\,s$^{-1}$, and a tangential peculiar velocity of 31.2$\pm$0.8\,km\,s$^{-1}$. The movement of V687~Car is shown in Figure~\ref{Figure:RGBY}~(top--right) and Figure~\ref{Figure:Measurement}, with the star shown to be moving in a north-westerly direction, aligned with the head-tail radio structure observed in the larger emission.

%The tangential \ac{PM} of V687~Car is 8.728\,mas\,yr$^{-1}$, with a tangential velocity of $\sim$77.4\,km\,s$^{-1}$, qualifying V687~Car as a possible high-velocity runaway star \citep{2022A&A...663A..39B}.

The projected \ac{PM} of V687~Car somewhat matches the geometry of the irregular elliptical shape of the radio--continuum emission, although the peculiar velocity indicates that it is moving more northern than north-westerly (See Figure~\ref{Figure:Measurement}). As it is possible that this is a chance alignment, we estimate the probability of a chance alignment using the \textit{Gaia} DR3 data. The probability of one or more stars with magnitude equivalent or brighter than V687~Car appearing in the Anglerfish emission area is given as $P(S>0)$, where $P$ is the Poisson distribution with rate $\lambda$, where $\lambda$ is the expected number of stars. We use the \textit{Gaia} DR3 G-band magnitude of 10.28, for this estimate. We note this magnitude is higher than the previous V-band magnitude of \citet{2007PZP.....7...20B}, possibly due to the different bandwidth. As this value is from the \textit{Gaia} catalogue, we use this for the probability calculation to ensure consistent cross-matching between the catalogue. We estimate the rate by measuring the number of candidate stars in a search area of radius 60\arcmin, with an area $\pi\times60^2 = 11309.73$ square arcminutes. 79 stars satisfying the criteria were found, giving a rate of $79/11309.73 = 0.0069$ stars per square arcminute. Using the Anglerfish elliptical area defined earlier as 4.72 square arcminutes, the expected rate is $\lambda=0.0069\times4.72 = 0.033$ stars per 4.72 square arcminutes. Therefore, the probability of one or more stars appearing in the region, independent of Anglerfish, is estimated as $P(S>0) = 0.0325$ using the Poisson distribution. This is indicative of a low probability of chance alignment.

Using a similar method explored in \citet{2025PASA...42..101B}, we are able to calculate V687~Car's trajectory and determine a theoretical age and expansion velocity of the supposed mass-loss shell. Figure~\ref{Figure:Measurement} shows measurements of the star's projected motion across the radio--continuum emission. Assuming that the centre of the ellipse used to measure the radio--continuum emission (See Section~\ref{results and discussion}) is close to the geometric centre of the shell, the point at which V687~Car comes closest to the centre of the emission is RA(J2000) $=$ 09$^{\rm h}$44$^{\rm m}$19.34$^{\rm s}$, Dec(J2000) $=-$75$^\circ$10$'$24.23$''$. The angular distance from V687~Car to the `centre' coordinates is $\sim$16$^{\prime\prime}$, and using the same method in \cite{2025PASA...42..101B}, we determine the time travelled to be 5060$\pm$130~yrs. Assuming that the mass-loss shell originated at this same point, we take the calculated time to be an estimate of the age of the shell. Measuring the longest distance of the travelled emission (90$^{\prime\prime}$, from the elliptical region), and dividing by the shell age, we determine an average expansion velocity of 175$\pm$5~km~s$^{-1}$. It is important to note that this assumes that the radio--continuum emission began at its apparent centre, and also assumes a 2-dimensional plane, making the age estimate a lower limit and the expansion velocity estimate an upper limit.

Considering our calculated average expansion velocity of 175$\pm$5~km~s$^{-1}$, it is important to note that the expected stellar wind velocity of an M-type giant does not typically exceed $\sim$20~km~s$^{-1}$ \citep{2015A&A...575A.105B,2017A&A...606A...6L}. This makes V687~Car an unlikely host for the radio emission, as the stellar wind output is not powerful enough to sustain a shell of this size. The spectral index range calculated suggests that the radio emission is consistent with both thermal and non-thermal origins (see Section~\ref{results and discussion}). The mass-loss episode scenario would be expected to generate thermal emission, which is consistent with the spectral index range.%, which would also argue against the mass-loss episode scenario if this is the case, as the radio emission would be expected to be thermal~\citep{1987LNP...291..322D,2002ApJ...580..459G}.

\subsection{Anglerfish as a High Latitude Supernova Remnant}
\label{SNR section}

We also consider the possibility that Anglerfish may be a high-latitude \ac{SNR}, similar to ones such as Calvera~\citep{2022A&A...667A..71A},  G70.0$-$21.5~\citep{2015ApJ...812...37F}, and G181.1+9.5~\citep{2017A&A...597A.116K}. The expected spectral index of an \ac{SNR} is $\sim-0.5$ (Galactic average is $-$0.51$\pm$0.01~\citep{2023ApJS..265...53R}), which is within the spectral index range given in Section~\ref{results and discussion}.
The spectral index is only weakly constrained however, and other physical properties act as a more accurate indicator for the Anglerfish's nature. At Anglerfish's direction ($l$\,=\,291.7\D, $b$\,=\,$-$16.5\D), we can calculate a maximum distance if we assume that it would be located within the Galactic disk. Assuming a maximum disk width of $\sim$1\,kpc we estimate a maximum likely distance of $D=1\textrm{~kpc}/\textrm{sin}(-16.5$\D$)=\,\sim3.5$\,kpc, corresponding to a physical diameter of $\sim$2\,pc. This would make Anglerfish one of the smallest \acp{SNR} discovered to date. There would only be one known \ac{SNR} with a smaller physical size, the SN~1987A with a diameter of 0.4\,pc, and one with a possibly similar size, the Galactic \ac{SNR} Perun, which may be as small as 2\,pc~\citep{2024MNRAS.534.2918S}. We note, however, that Perun's size was not fully constrained, and this smallest size is the lower end of a given diameter range due to complications in measuring an exact distance. If we instead assume that Anglerfish may be located outside of the Galactic disk, at a latitude of $-$16.5\D, the Milky Way extends to a maximum distance of $\sim$20\,kpc~\citep[][their Figure~16]{Churchwell2009}. This maximum distance thus corresponds to a maximum physical size of $\sim$10\,pc diameter. This is within the Galactic \ac{SNR} population, but is relatively small compared to the Galactic average~\citep[30.5$\pm$1.7\,pc; ][]{2023ApJS..265...53R}.

Another issue with this scenario is the radio surface brightness. Smaller \acp{SNR} are expected to be younger and thus have a higher radio surface brightness. Using the measured flux density at 944\,MHz and the estimated spectral index range, we calculate surface brightness as $\Sigma_{\textrm{1\,GHz}}=S_{\textrm{1\,GHz}}/\Omega$, where $S_{\textrm{1\,GHz}}$ is the flux density scaled to 1\,GHz and $\Omega$ is the calculated surface area in steradians using a measured radius of $r$\,=\,55\arcsec. As the flux density scaling uses the estimated spectral index, we use the upper and lower limits to calculate two surface brightness values, $\Sigma\,=\,2.2\times10^{-22}$\,W\,m$^{-2}$\,Hz$^{-1}$ for $\alpha\,=\,-0.6$ and $\Sigma\,=\,2.1\times10^{-22}$\,W\,m$^{-2}$\,Hz$^{-1}$ for $\alpha\,=\,-1.2$. We note that we have not included the uncertainties in the spectral indices in this calculation. This is primarily because the surface brightness values are used to estimate an empirical relationship, and a small change in spectral index will not substantially alter them. We use this value to place Anglerfish in the context of the Galactic \ac{SNR} population using the established statistical $\Sigma-D$ relation ~\citep{Pavlovic2018}. This is a statistical, empirical relationship which compares an \ac{SNR}'s radio surface brightness with its physical size and has been used to analyse the Galactic \ac{SNR} population. Most \acp{SNR} follow a typical trend where the surface brightness decreases by size, and the Galactic population is mostly located in a particular region of the $\Sigma-D$ diagram. We assume a most likely diameter of $\sim$2\,pc for Anglerfish if it is a Galactic object and compare it with the Galactic population distribution ~\citep[][their Figure~3]{Pavlovic2018}. We find that these values would place Anglerfish in the lower left part of this graph, well outside of the main Galactic \ac{SNR} population. A maximum diameter of 10\,pc would be more likely for the \ac{SNR} scenario, and would place Anglerfish closer to the Galactic \ac{SNR} population, but it would still be an outlier. While this is an empirical relationship and there are known \acp{SNR} that are outliers to this population, such a large difference argues against the \ac{SNR} scenario.

Additionally, \acp{SNR} are typically detected at other wavelengths, such as optical or X-ray, which is not the case for Anglerfish. In particular, for the emission to be Galactic, it must be quite small, meaning that it is more likely to be a younger \ac{SNR}, which are typically expected to have brighter X-ray emission. Overall, the contradictions in size, distance, and brightness, and the lack of detection at other frequencies, make the \ac{SNR} interpretation unlikely.

\subsection{Anglerfish as an ORC Candidate}
\label{ORC candidate}

We also consider the possibility that Anglerfish is a type of celestial object known as an \ac{ORC}. \acp{ORC} generally share a set of common properties required for the classification; they are typically centred on massive elliptical galaxies, they exhibit physical sizes of a few hundred kpc, and the diffuse component is seen exclusively at radio frequencies \citep{Norris2021ORC,Gupta2022,2025arXiv250904981T}. The most common radio morphology consists of edge-brightened, near-circular emission. There are some variations within the known \ac{ORC} population for certain properties. Some \acp{ORC} show more complex internal ring-like structures~\citep{2022MNRAS.513.1300N}, and some display additional structure adjacent to the main circular structure  (e.g. ORC1; \citealt{2022MNRAS.513.1300N}). Additionally, some display a double structure consisting of intersecting rings~\citep{2022MNRAS.513.1300N,2024A&A...686A..44R,2025MNRAS.543.1048H,2025arXiv250904981T}, which may also be the case for Anglerfish. These double structures are more likely to form from a dynamic origin, as this would likely be required to form large, several-hundred-kpc size rings of equal size on either side of a galaxy.

Several of these properties are also shared by Anglerfish, where we see a structure of exclusively radio-continuum emission with an optical/IR source near the geometric centre.  This \ac{IR} source is identified as \wisegal\ in the catalogue of \citet{2013wise.rept....1C}.
%The galaxy has catalogued \ac{IR} fluxes and a listed redshift value of $z=0.0704\pm 0.0410$ \citep{2023A&A...674A...1G}. 
This source is located near the geometric centre, slightly offset towards the head of the structure (Figure~\ref{Figure:RGBY}). Due to this location, we investigate it as a host galaxy for the Anglerfish emission in the context of a possible \ac{ORC} scenario.

We use the \ac{IR} \ac{WISE} observations of \cite{2013wise.rept....1C} to measure the W1$-$W2 and W2$-$W3 colours of \wisegal\ to help classify the galaxy. The recorded \ac{WISE} magnitudes are W1 (3.4\,$\mu$m) = 10.81$\pm$0.02 mag, W2 (4.6\,$\mu$m) = 10.54$\pm$0.02 mag, and W3 (12\,$\mu$m) = 9.58$\pm$0.03 mag~\citep{2013wise.rept....1C}, giving values of W1$-$W2 = 0.267$\pm$0.04 mag and W2$-$W3 = 0.960$\pm$0.06 mag. We compare these values with the colour-colour plot of \citet[their Figure~12]{2010AJ....140.1868W}, which places \wisegal\ in the elliptical galaxy region. Due to potential \ac{WISE} photometry contamination issues from the nearby star V687~Car listed in the catalogue, we further check this classification using the observed \textit{Gaia} DR3 colours and calculate the colours G$-$RP = 3.9 and BP$-$G = --2.3. These values place \wisegal\ in the galaxy section of the \textit{Gaia} colour-colour plot of \citet{2024A&A...692A.154W}. This makes \wisegal\ a potential host galaxy for Anglerfish if it is an \ac{ORC}.

\wisegal\ has a literature redshift of $z_{\rm UGC}=0.0704\pm 0.0410$. This was determined using \textit{Gaia} \ac{DR3} as described by \citet{2023A&A...674A..41G}, which contains a catalogue of calculated redshifts for galaxies (i.e. the \ac{UGC} Catalogue) observed with \textit{Gaia}, using a \ac{SVM}, based on the RP and BP low resolution spectra. At the \textit{Gaia} $z_{\rm UGC}$, using the radius of the measured circular component of the radio emission (55$^{\prime\prime}$), we calculate a linear diameter of 153$\pm$82\,kpc, assuming the cosmological parameters $H_0\,=\,67.31\,$\kms, $\Omega_M=0.315$, and $\Omega_L=0.685$~\citep{2020A&A...641A...6P}. This size estimate is somewhat smaller than typical for an \ac{ORC} \citep{2022MNRAS.513.1300N}. 

We note that the redshifts calculated for the UGC within \textit{Gaia} are estimated via the \ac{SVM} machine learning algorithm, using very low-resolution spectra across a very narrow wavelength, with a maximum redshift of $z < 0.6$. Further, there is at least 2\% of the $\sim248\mathrm{k}$ sources with a spectroscopically measured redshift that are incorrectly estimated --- i.e. for the majority of sources, the \textit{Gaia} estimated redshift is likely to be accurate \citep{2023A&A...674A..31D}, but where there is additional photometry available, additional estimates may prove more accurate. Given we have access to broadband photometry across a much wider band, from optical to infrared, we also calculate a photometric redshift using the SkyMapper Southern Sky Survey \citep{2007PASA...24....1K} and AllWISE photometry. We use the k-Nearest Neighbours machine learning method described by \citet{2022A&C....3900557L, 2023PASA...40...39L}.
We use the DR4 \citep{2024PASA...41...61O} values, which give an r-mag of 15.3, and we get a redshift of $z_{\rm ph}$=0.65$\pm$0.10. This corresponds to a diameter of 788\,kpc for Anglerfish (using the same cosmological parameters as above), more typical for \acp{ORC} \citep{2022MNRAS.513.1300N}.

We also estimate the stellar mass using the \ac{WISE} values~\citep{2013wise.rept....1C}, and applying the K-corrections from \citet{2010ApJ...713..970A} to obtain rest-frame magnitudes at both possible redshift values. Using the stellar mass-to-light ratio from \citet{2017ApJ...836..182J}, we estimate stellar masses of $M_{\odot}$=2.6$\times$10$^{10}$ for $z_{\rm sp}$=0.0704 and $M_{\odot}$=2.2$\times$10$^{10}$ solar masses for $z_{\rm ph}$=0.65.
\acp{ORC} are also characterised by having average spectral indices of $\alpha\sim-1$~\citep{Norris2021ORC}, which is within with the range calculated in Section~\ref{results and discussion}.

Morphologically, Anglerfish shares some characteristics with typical \acp{ORC}, but there are some slightly differing properties that must be discussed. Anglerfish is not as obviously limb-brightened as some other \acp{ORC}, e.g. ORC1 \citep[also known as ORC~J2103$-$6200; ][]{2022MNRAS.513.1300N}, and there is a patch of emission extending out of the south-eastern side, making the shape not perfectly circular. There are known \acp{ORC} which display similar additional structure, and so such asymmetry does not preclude an \ac{ORC} classification. There is the possibility that the extended emission may be unrelated to the circular emission, and thus the circular region would resemble a more typical \ac{ORC} structure~\citep{Norris2021ORC,10.1093/mnrasl/slab041,Norris2025}. This scenario is highly unlikely however, as it would require a chance coincidence of two overlapping regions of diffuse radio emission. As the Anglerfish is located at a high Galactic latitude, this coincidence would be very unlikely, and the entire emission is likely part of the same physical structure.

If the extended emission is associated and Anglerfish is not circular, this does not preclude an \ac{ORC} classification. While \acp{ORC} are typically circular, there are some observed that deviate from this perfect symmetry. For example, ORC~1 shows a generally circular structure, but when observed in more detail, some asymmetry becomes visible in the circular shape~\citep{Norris2021ORC}. There are some extensions of emission, particularly on the north-western edge, which deviate from this symmetry and make a more elliptical shape. This is less pronounced than in the case of Anglerfish, however. Another possible extension is faintly visible in ORC~4. It has been suggested that this extension may be caused by an orientation effect of a double-ring structure, where one ring is appearing behind the other, with the orientation angle causing a slight offset~\citep{Norris2021ORC}. Therefore, slightly asymmetric morphologies are not unheard of in the \ac{ORC} class, but without a definite origin for the \ac{ORC} phenomenon, it is difficult to state if it is an expected property. %Conversely, it is also possible that Anglerfish's unusual morphology may be due to a chance coincidence with unrelated emission.}

Multiple origin scenarios for \acp{ORC} are discussed in the literature, including, but not limited to, \ac{SMBH} merger events, galaxy mergers, and remnant lobes from radio galaxies \citep{2022MNRAS.513.1300N, Dolag2023, Shabala2024}. Some asymmetry is predicted in some of these origin formation scenarios; for example, the phoenix origin hypothesis for \acp{ORC}~\citep[][see their Figure~2]{Shabala2024}. This scenario posits that \acp{ORC} are remnant lobes from powerful radio galaxies which have been re-energised by the passage of energetic shocks. The simulations involved in this scenario predict that some \acp{ORC} may show an incomplete structure with offset extensions if the shocks are viewed from different angles. %They also predict that this can cause the central galaxy to be offset slightly from the geometric centre. 
For example, see the model of an \ac{ORC} with a 75\D\ viewing angle and a 400\,Myr age as presented in \citet[][see their Figure~A2]{Shabala2024}. For this origin scenario, it is also expected that there may be X-ray shocks observable near the \ac{ORC}, and perhaps an observable X-ray cavity associated with a secondary (invisible in the radio), radio lobe. Due to this, high-resolution X-ray observations, such as by the {\it Chandra} or {\it XMM-Newton} telescopes, would be useful to better determine between these scenarios. Additionally, a better constraint on the radio spectral index, as well as high-resolution observations at other frequencies, such as optical, would be useful in more definitively ruling out multi-frequency counterparts, thus arguing for the \ac{ORC} scenario.

We also note that the possible host galaxy \wisegal\ is not located exactly at the geometric centre of the emission, but is slightly offset to the north-west. If we take the centre of the circular emission as defined in Section~\ref{results and discussion}, we find that \wisegal\ is offset by 12\farcs4 from the centre. Slight offsets of the host galaxy from the observed geometric centre have been observed in other \acp{ORC}~\citep[e.g ORC J0219$-$0505][]{Norris2025}. The host offset observed for ORC J0219$-$0505 is 4\arcsec\ ($\sim13$\,kpc at their measured redshift of $z\,=\,0.196$), significantly less than that observed for Anglerfish in terms of absolute size (12.4\arcsec, $\sim$87\,kpc at $z\,=\,0.65$). An interesting scenario, however, is that of the phoenix hypothesis discussed by \citet{Shabala2024}, who predict a host offset in their scenarios, and map these values as a percentage of host offset divided by major axis. If we calculate the offset/major axis for Anglerfish, we get values of 0.225 (if we use the 55\arcsec\ radius circular region) and 0.138 (if we use the larger elliptical region as defined in \ref{results and discussion}). A similar calculation using the values for ORC J0219$-$0505 gives a value of 0.229, quite similar to that of the Anglerfish circular calculation. Therefore, this offset does not preclude an \ac{ORC} candidate classification, particularly in the case of the phoenix origin scenario, and if Anglerfish is identified as an \ac{ORC}, it could help provide evidence for the origin of these objects.

There is also the possibility that Anglerfish may be an object known as \acp{GLARE}~\citep{2025arXiv250608439G}. It has been suggested that \acp{GLARE} may be \ac{ORC} precursors, or \acp{ORC} at a different evolutionary stage, and these objects can display more irregularly shaped emission than typical \acp{ORC}. If Anglerfish were a \ac{GLARE}, then it may be a type of ``rectangular \ac{GLARE}'', following the classification scheme of \citet{2025arXiv250608439G}. 

It is also possible that the galaxy \wisegal\ is a chance alignment with the emission, and we investigate this scenario in a similar way as done in section~\ref{stellar mass loss}. Using the same 1\D\ radius search region as used for the V687~Car calculation, we find 84 catalogued galaxies in the region from the \ac{NED} database. This gives a rate of 0.0074 galaxies per square arcminute, and multiplying this by the Anglerfish area of 4.72 square arcminutes, we find an expected rate of 0.035 galaxies per 4.72 square arcminutes. Therefore, the probability of one or more galaxies appearing in this region, independent of Anglerfish, is estimated as $P(G>0) = 0.0344$ using the Poisson distribution. This indicates a low probability of a chance alignment, similar to the case for V687~Car. Therefore, it is not possible to statistically determine whether one of the objects is more likely to be associated with the emission; thus, this alignment is not a good discriminator among the possible scenarios.

\section{Conclusion}
\label{Section:Conclusion}

We report the detection of a radio-continuum source observed with \ac{ASKAP}, which we name Anglerfish -- EMU~J094412$-$751016. We discuss three main origin scenarios: the first is a mass loss episode from the variable M-Type giant star V687 Car, the second is a Galactic \ac{SNR}, and the third is an \ac{ORC} candidate with an elliptical host galaxy, \wisegal. 

Given the star's weak winds compared to the calculated theoretical expansion velocity, we deem it unlikely that V687~Car is physically associated with the radio-continuum emission and conclude that it is likely a chance superposition.

We also investigate whether Anglerfish may be a high-latitude Galactic \ac{SNR}, but find that the observed size and brightness are not consistent with the Galactic \ac{SNR} population. For Anglerfish to be located within the \ac{MW}, we find that its size would have to be smaller than expected for a Galactic \ac{SNR}, and it would lie well outside of the Galactic $\Sigma-D$ distribution. We therefore deem this scenario unlikely without more compelling supporting evidence.

Finally, we present the \ac{ORC} scenario as the most likely scenario. We find the \ac{IR} source \wisegal\ is a promising candidate for a host galaxy, as the \ac{IR} colours indicate that it is an elliptical galaxy centred on the radio emission. The distance to this galaxy would give Anglerfish a physical size of several hundred kiloparsecs, depending on the redshift used, qualifying it as a possible \ac{ORC} candidate. Similarly, the estimated spectral index range is consistent with an \ac{ORC} scenario.

Overall, we determine the \ac{ORC} scenario to be most likely with the current available data, and propose this object as an \ac{ORC} candidate. A definitive classification is not currently possible, however, and future multi-frequency observations, particularly in the radio, X-ray, and optical regimes, are vital to better constrain Anglerfish's properties and determine its nature.

\begin{acknowledgement}

This scientific work uses data obtained from Inyarrimanha Ilgari Bundara, the CSIRO Murchison Radio-astronomy Observatory. We acknowledge the Wajarri Yamaji People as the Traditional Owners and native title holders of the Observatory site. CSIRO’s ASKAP radio telescope is part of the Australia Telescope National Facility (\url{https://ror.org/05qajvd42}). Operation of ASKAP is funded by the Australian Government with support from the National Collaborative Research Infrastructure Strategy. ASKAP uses the resources of the Pawsey Supercomputing Research Centre. Establishment of ASKAP, Inyarrimanha Ilgari Bundara, the CSIRO Murchison Radio-astronomy Observatory and the Pawsey Supercomputing Research Centre are initiatives of the Australian Government, with support from the Government of Western Australia and the Science and Industry Endowment Fund.

This work has made use of data from the European Space Agency (ESA) mission {\it Gaia} (\url{https://www.cosmos.esa.int/gaia}), processed by the {\it Gaia} Data Processing and Analysis Consortium (DPAC, \url{https://www.cosmos.esa.int/web/gaia/dpac/consortium}). Funding for the DPAC has been provided by national institutions, in particular the institutions participating in the {\it Gaia} Multilateral Agreement.

\end{acknowledgement}

\paragraph{Funding Statement:}
%MDF and GR acknowledge \ac{ARC} funding through grant DP200100784.  
DU acknowledge the financial support provided by the Ministry of Science, Technological Development and Innovation of the Republic of Serbia through the contract 451-03-66/2024-03/200104 and for support through the joint project of the Serbian Academy of Sciences and Arts and Bulgarian Academy of Sciences ``Optical search for Galactic and extragalactic \acp{SNR}''. CJR acknowledges financial support from the German Science Foundation DFG, via the Collaborative Research Center SFB1491 ``Cosmic Interacting Matters – From Source to Signal''.

\paragraph{Data Availability Statement:}

The \ac{ASKAP} data used in this article are available through the \ac{CASDA} (\url{https://research.csiro.au/casda}).

%\endnote in some journals will behave like \footnote; and \printendnotes will not output anything. 
\printendnotes

\bibliography{Anglerfish}
%\printbibliography

\appendix
\renewcommand{\thefigure}{A\arabic{figure}}
\setcounter{figure}{0}

\begin{figure*}
\centering
    \includegraphics[width=\textwidth]{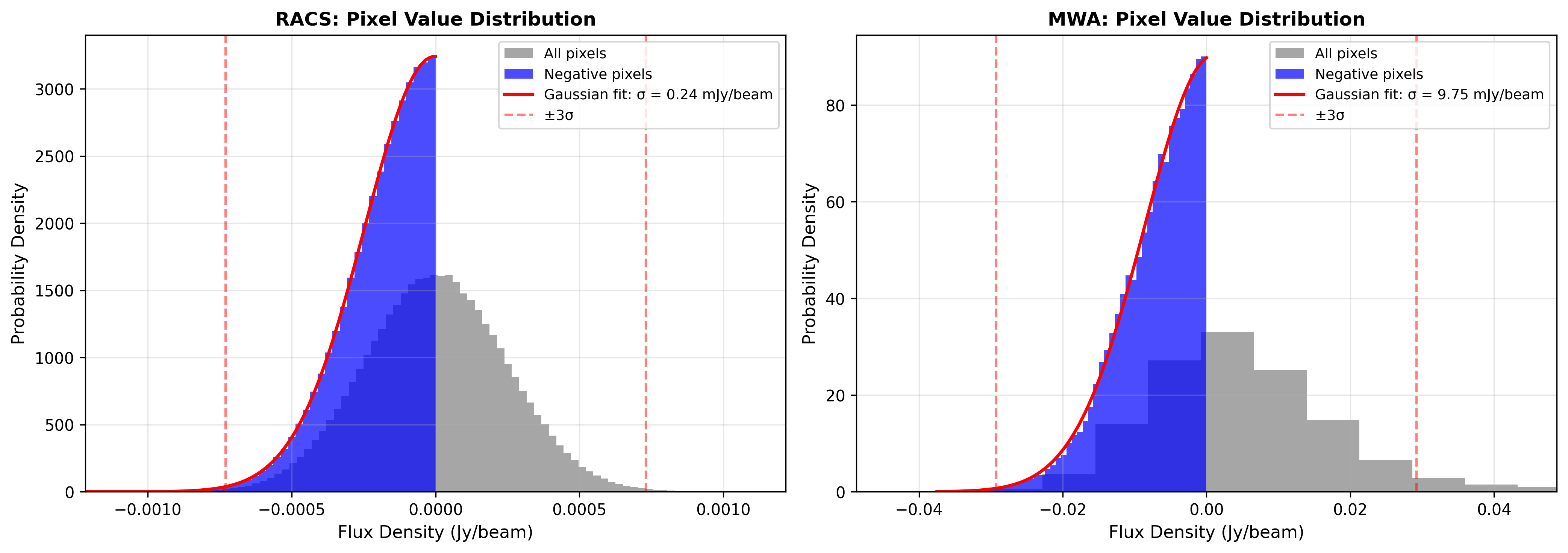}
    \caption{Histogram of the pixel values for \ac{RACS} image (left) and \ac{MWA} image (right). All pixel values are shown in grey, and the negative pixel values are shown in blue, both normalised to unit area. The different heights of the grey and blue histograms result from their different sample sizes when normalised to probability density. The blue histogram represents only the negative pixels (a narrower range) while the grey inclues all pixels (spanning a wider range including positive sources). The fitted Gaussian (the thick red line) was fit to only the negative values to estimate the background noise level, and the dashed red vertical lines on the left panel show the $\pm$3$\sigma$ limits.} 
    \label{fig:Noise}
\end{figure*}

\end{document}